\documentclass[12pt]{article}

\usepackage{scicite}

\usepackage{times}

\usepackage{lineno}

\usepackage{graphicx}

\usepackage[version=4]{mhchem} 

\usepackage{rotating} 

\usepackage{newtxtext,newtxmath}

\usepackage[doublespacing]{setspace}

\begin{document} 

\title{A coupled model of episodic warming, oxidation and geochemical transitions on early Mars} 

\author
{Robin~Wordsworth,$^{1,2}$ Andrew~H.~Knoll,$^{3}$ Joel~Hurowitz,$^{4}$\\
 Mark~Baum,$^{2}$ Bethany~L.~Ehlmann,$^{5,6}$ James~W.~Head,$^{7}$ Kathryn~Steakley,$^{8}$\\
\normalsize{$^{1}$School of Engineering and Applied Sciences, Harvard University, Cambridge MA 02137, USA}\\
\normalsize{$^{2}$Department of Earth and Planetary Sciences, Harvard University, Cambridge MA 02137, USA}\\
\normalsize{$^{3}$Department of Organismic and Evolutionary Biology,} \\\normalsize{ Harvard University, Cambridge MA 02137, USA}\\
\normalsize{$^{4}$Department of Geosciences, Stony Brook University, Stony Brook NY 11794-2100, USA}\\
\normalsize{$^{5}$Division of Geological and Planetary Sciences,} \\\normalsize{California Institute of Technology, Pasadena CA 91125, USA}\\
\normalsize{$^{6}$NASA Jet Propulsion Laboratory, Pasadena CA 91109, USA}\\
\normalsize{$^{7}$Department of Earth, Environmental and Planetary Sciences,}\\\normalsize{ Brown University, Providence, RI 02912, USA}\\
\normalsize{$^{8}$NASA Ames Research Center, Moffett Blvd, Mountain View, CA 94035, USA}\\
\\
}
\date{}

\maketitle

\newpage

\textbf{Reconciling the geology of Mars with models of atmospheric evolution remains a major challenge. Martian geology is characterized by past evidence for episodic surface liquid water, and geochemistry indicating a slow and intermittent transition from wetter to drier and more oxidizing surface conditions. Here we present a new model incorporating randomized injection of reducing greenhouse gases and oxidation due to hydrogen escape, to investigate the conditions responsible for these diverse observations. We find that Mars could have transitioned repeatedly from reducing (\ce{H2}-rich) to oxidizing (\ce{O2}-rich) atmospheric conditions in its early history. Our model predicts a generally cold early Mars, with mean annual temperatures below 240~K. If peak reducing gas release rates and background \ce{CO2} levels are high enough, it nonetheless exhibits episodic warm intervals sufficient to degrade crater walls, form valley networks and create other fluvial/lacustrine features. Our model also predicts transient buildup of atmospheric \ce{O2}, which can help explain the occurrence of oxidized mineral species such as manganese oxides at Gale Crater. We suggest that the apparent Noachian--Hesperian transition from phyllosilicate deposition to sulfate deposition around 3.5 billion years ago can be explained as a combined outcome of increasing planetary oxidation, decreasing groundwater availability and a waning bolide impactor flux, which dramatically slowed the remobilization and thermochemical destruction of surface sulfates. Ultimately, rapid and repeated variations in Mars' early climate and surface chemistry would have presented both challenges and opportunities for any emergent microbial life.}

Today, Mars is dry, cold and volcanically quiet, with mean surface temperatures of around 210~K and a thin ($\sim 6$~mbar) \ce{CO2}-dominated atmosphere. Extensive geomorphological and geochemical evidence indicates, however, that in the distant past, surface conditions on Mars were dramatically different, with enhanced fluvial erosion, aqueous geochemical alteration and sediment deposition \cite{Carr2010,Grotzinger2014}. The most plausible explanation for the aqueous alteration of Mars' surface is strong greenhouse warming from a thicker early atmosphere, although details of the warming mechanism continue to be debated [e.g., \cite{Halevy2007,Tian2010,Toon2010,Forget2013,Haberle2017,Ramirez2018}]. Geomorphic and geochemical analyses suggest that in total, between around $10^4$ and $10^7$ years of warm conditions were required to erode observed valley networks, deposit sediment in craters and form Al-/Fe- phyllosilicate weathering sequences \cite{Toon2010,Grotzinger2014,Hoke2011,bishop2018surface,Lapotre2020}. In addition, abundant ancient exposures of unaltered igneous minerals such as olivine and the relative absence of surface carbonates on Mars indicate that large bodies of surface liquid water were likely not present over periods much longer than a few million years \cite{olsen2007,Ehlmann2014,Niles2013}.

The chemistry of the martian surface has also varied over time. Orbital and rover observations of surface mineralogy indicate extensive phyllosilicate formation at the Noachian surface and/or in the crust \cite{Ehlmann2011} and an increase in the deposition of sulfates in the Hesperian \cite{Bibring2006}. Martian meteorite data suggest that much of Mars' mantle is more reducing than Earth's \cite{Wadhwa2001}, and hence that its early atmosphere was also at least intermittently reducing \cite{Ramirez2014,Wordsworth2017}. Some Mars meteorites also preserve non-zero sulfur mass-independent fractionation (MIF) signatures, suggesting sulfur deposition during anoxic atmospheric intervals \cite{Lorand2020}. Today, however, Mars' surface is highly oxidized, with abundant ferric iron lending the planet its reddish appearance and species such as hydrogen peroxide and perchlorates present in the atmosphere and regolith. 
Rover observations at both Meridiani Planum and Gale Crater have also shown strong variability in mineral redox chemistry \cite{Hurowitz2017}, with oxidizing surface conditions around the Noachian-Hesperian boundary ca. 3.5~Gya (Figure~1) suggested by the presence of concentrated hematite and manganese oxide \cite{Lanza2016,Hurowitz2017,Hurowitz2010}. The manganese deposits in particular appear to require the simultaneous presence of liquid water and strong oxidants such as \ce{O2} or ultraviolet radiation \cite{Lanza2016}. Formation of manganese oxide from manganese carbonate via ultraviolet photo-oxidation may be possible in certain circumstances \cite{Liu2020}, but it requires the prior presence of rhodochrosite deposits  and has difficulty explaining manganese oxide formation in sub-surface fracture fills \cite{Lanza2016}.

\subsection*{A coupled atmospheric evolution and climate model}

To investigate how these diverse, apparently contradictory strands of geologic evidence can be reconciled, we have constructed an integrated atmospheric evolution and climate model. A key feature of our model is that it is stochastically forced in a way that captures the episodic nature of the key processes that altered Mars' atmosphere in its early history. We represent the release of reducing gases to the atmosphere due to meteorite impacts \cite{Haberle2019}, volcanism \cite{Ramirez2014} and crustal alteration \cite{Hurowitz2010,Wordsworth2017,Chassefiere2016,tosca2018magnetite} in a generalized way by randomly sampling from a power law distribution, which has probability density function
\begin{equation}
p(x) = Cx^\gamma,  \label{eq:probeqn}
\end{equation}
where $x$ is the reducing gas input rate in mol/s defined on the interval $[x_0,x_1]$ and $\gamma$ is a power-law coefficient  \cite{methods}. We assume a default value of $\gamma=-1.95$ based on observed meteorite impact and volcanic outgassing distributions from crater-counting and terrestrial occurrence data, respectively, but tested the sensitivity of the results to variations in this parameter \cite{methods}. $C$ is a normalization constant defined such that 
\begin{equation}
\int_{x_0}^{x_1} p(x) dx = 1. \label{eq:probnorm}
\end{equation}
In our model, the distribution mean $\mu$ and the maximum input value $x_1$ are used to determine the minimum input value $x_0$, which in turn determines $C$ and hence the probability density function $p(x)$ \cite{methods}. We allow $\mu$ to decrease slowly with time to represent the secular decay in all input processes over martian history. Escape of hydrogen and oxygen to space via diffusion-limited and non-thermal escape processes \cite{jakosky2018loss} is also included (Fig.~S1), as is oxidative weathering of the crust. Changes in surface temperature over time due to greenhouse warming by \ce{CO2}, \ce{H2O} and reducing gases are calculated using a line-by-line radiative-convective model {that includes collision-induced absorption (CIA) effects} \cite{methods}. \ce{CO2} evolution through time is represented by an empirical function constrained by observations and modeling, although we study model sensitivity to variations in total \ce{CO2}. To remain conservative, we take \ce{CO2-H2} collision-induced absorption into account using recent experimental results \cite{turbet2020measurements} rather than slightly stronger previous theoretical estimates \cite{Wordsworth2017}.

Figure~\ref{fig:coupled} shows the model output for an example run with time mean reducing gas input $\overline \mu = 1.5\times10^4$~mol/s and variability ratio $\beta \equiv x_1 / \overline \mu = 10^3$. As can be seen, both the atmospheric redox state (Fig.~1B) and the surface temperature (Fig.~1C) are highly variable in time, showing rapid and repeated fluctuations between {warmer, more reducing and colder, more oxidizing conditions} during Mars' early history. The bulk oxidation state of the planet increases monotonically with time due to non-stochiometric hydrogen escape to space \cite{methods}, but the atmosphere undergoes multiple reducing episodes in both the Noachian (reducing around 13\% of the time) and the Hesperian / Amazonian (reducing around 6\% of the time) (see also Figs.~S2 and S3).  The time-mean temperature is far below the melting point of liquid water, but the total integrated duration of warm periods is consistent with the timescale required from geological constraints (Fig.~1D).  

Figure~\ref{fig:pspace_span} summarizes the results of varying both $\overline \mu$ and $\beta$, with additional runs varying other parameters shown in Fig.~S4. For each case, we show the mean and standard deviation of an ensemble of 256 stochastic simulations. When variability is high, our model produces integrated warm periods sufficient to match geomorphic observations for mean rates of $4\times10^3$ to about $1\times10^5$~mol/s of hydrogen input. 
These mean rates are in the range of upper limits from volcanism ($1.2\times10^4$~mol/s), although obtaining the extreme levels of episodicity in volcanic activity required is challenging \cite{methods}. They are also within the range of bolide impact upper limits. The impact basins large enough to drive extended warming episodes alone (e.g. Hellas, Argyre and Isidis) can plausibly explain much of the Noachian-era fluvial crater degradation \cite{mangold2012chrono,Haberle2019}, although they appear 100 to 500~My older than many valley networks based on crater statistics \cite{fassett2008sequence}. Finally, upper limits on H release from crustal alteration ($1.5\times10^4$~mol/s) also match the required constraints, and crustal alteration processes may have exhibited substantial episodicity via nonlinear aqueous feedbacks and coupling to impact thermal forcing \cite{methods,tosca2018magnetite}. Additional effects such as warming from \ce{C_nH_n} hydrocarbons or cirrus clouds produced following \ce{CH4} photolysis \cite{Wordsworth2017}, and a higher solar flux due to uncertainties in reported terrain ages and early solar mass loss are neglected here but would allow the model to match observations at lower $\beta$ and $\overline \mu$ values \cite{methods}. When the average model H input flux is low, build-up of \ce{O2} in Mars' atmosphere continues to the present day, in conflict with observations. Average H input fluxes above $2 \times 10^6$~mol/s, which are required for permanently `warm and wet' early conditions, are not plausible by any known mechanism and produce permanently reducing surface conditions that are hard to reconcile with the observed geochemical record. 

\subsection*{Implications for Mars' sedimentary geochemistry}

As well as providing a way to explain many features of Noachian to early Hesperian fluvial and lacustrine geomorphology, intermittent warming by reducing gases on a planet whose surface oxidizes over time can help account for key features of Mars' aqueous mineralogy. Warming of up to a few million years total duration is consistent with the formation timescales of Noachian/Early Hesperian Al/Fe phyllosilicate weathering sequences \cite{bishop2018surface} and the lack of evidence for open-system, weathering-driven chemical fractionation of ancient terrains \cite{Ehlmann2011}. Assuming cold-trapping of \ce{H2O} on the southern highlands due to adiabatic cooling \cite{Wordsworth2013a}, episodic warming (with subsequent localized hydrological activity and evaporation) is also consistent with the formation of chloride deposits on the Noachian southern highlands \cite{Osterloo2010,hynek2015late}.   

One particularly visible and puzzling feature of martian aqueous mineralogy is the broad trend from phyllosilicate formation in the Noachian to mainly evaporite formation, including regional sulfate deposition, in the subsequent Hesperian \cite{Bibring2006,Halevy2007,mclennan2005provenance,Ehlmann2014}. Previously, this transition has been interpreted in terms of a change in the rate or style of volcanic outgassing at the Noachian-Hesperian boundary \cite{Bibring2006,Halevy2014,gaillard2013geochemical}. While this may have been a contributing factor, overall there was more volcanic outgassing in the Noachian than in the Hesperian, and changes in the quantity of S outgassed as a function of atmospheric pressure and melt composition are relatively modest over plausible ranges \cite{gaillard2013geochemical}. The detected sulfates on Mars probably represent only a fraction of the sulfur outgassed throughout martian history \cite{gaillard2013geochemical}, which  suggests that a larger inventory of S may currently be present near the surface or deeper in the crust, given Mars' modest degree of crustal recycling.

We hypothesize that the increase in sulfate deposition from the Noachian into the Hesperian was directly linked to the secular changes in the planet's redox state through time, for three reasons. First, oxidation of the upper martian mantle due to metamorphism of hydrated crust \cite{wade2017divergent} could potentially have moderately increased the fraction of S outgassed per unit volume of melt by the Hesperian \cite{gaillard2013geochemical}. Second, under the fluctuating wet-dry climate conditions driven by intermittent reducing gas fluxes, evaporite formation would have been ongoing, along with clay formation, throughout the Noachian. However, the preservation of such deposits against later remobilization would have only been favored toward the waning stages, from 3.5~Ga onward \cite{Milliken2009,Zolotov2016}, which is consistent with the geologic record of chlorides and sulfates \cite{Ehlmann2014}. Third, sulfate minerals can be converted to minerals like pyrite when dissolved under reducing conditions.  

As can be seen from Figure~\ref{fig:S_thermochem}a, for \ce{O2} fugacities below around $10^{-38}$~atm, pyrite and pyrrhotite are predicted to be the primary precipitating minerals from an S-rich solution across a wide range of pH conditions. While {most} outgassed \ce{SO2} and \ce{H2S} is converted efficiently to atmospheric \ce{H2SO4} aerosols and hence ultimately to surface sulfate deposits even in an \ce{H2}-rich atmosphere \cite{sholes2017anoxic}, aqueous thermochemical sulfate reduction (TSR) {can proceed slowly in warm environments and } becomes rapid once temperatures exceed $\sim 150^\circ$C. The frequency of large bolide impacts, {high geothermal heat flux} and relative abundance of water during the pre- to mid-Noachian means that {moderate to} high temperature aqueous environments may have occurred repeatedly in combination with reducing atmospheric conditions early in martian history \cite{Haberle2019,Segura2002,steakley2019testing}, resulting in the repeated {dissolution and reduction} of early surface sulfate deposits. Interestingly, TSR of surface-derived sulfates has very recently been proposed as an explanation for non-zero S-MIF values in the pyrite of martian meteorite NWA 7533 \cite{Lorand2020}. 

In the short-term aftermath of a large impact (timescales of months to years), hot ejecta causes a thermal wave to propagate downwards into the regolith and widespread rainfall occurs \cite{Segura2002,steakley2019testing}, leading to high-temperature aqueous chemical alteration on regional to global scales \cite{methods,Segura2002,steakley2019testing,palumbo2018impact,turbet2020environmental}. Figure~\ref{fig:S_thermochem}b shows the amount of sulfate minerals on the surface that would have been thermochemically reduced due to this effect as a function of ejecta layer depth. Following the largest impactors such as Isidis and Hellas, up to several hundred meters of the crust over wide regions of the surface could have undergone rapid sulfate reduction via this effect. Longer term sub-surface alteration at more moderate temperatures, while hard to quantify in the Noachian, could have contributed to TSR further. Hydrological and thermochemical suppression of sulfate preservation until the late Noachian can therefore help to explain the paucity of evaporite preservation in the Noachian record.

Finally, redox fluctuations can also help explain the appearance of hematite and manganese oxide deposits in sediments at Meridiani Planum, and at Gale Crater, where they are frequently found in combination with more reducing species \cite{Hurowitz2017}.  Our model predicts intervals of elevated atmospheric oxygen (\ce{O2} {levels up to around 0.05~bar in our nominal simulations; Fig.~\ref{fig:coupled}}B) throughout the Noachian and Hesperian. In these intervals, upper layers of the regolith would slowly oxidize via dissolution of \ce{O2} into lava, dry weathering reactions, formation of oxychlorine species \cite{mitra2020capacity}, and adsorption of volatile species like \ce{H2O2}. During warming intervals, some of this regolith would have been transported as sediment to standing bodies of water. The extremely slow rate of reaction of \ce{H2_{(aq)}} with oxidized species like \ce{Fe^{3+}}, combined with photo-oxidation and other processes that produce \ce{H2_{(aq)}} by oxidizing \ce{Fe^{2+}} to \ce{Fe^{3+}} \cite{Hurowitz2010,tosca2018magnetite}, mean these sediments would have remained out of chemical equilibrium with the atmosphere, plausibly leading to the local formation within the sediments of oxidized minerals such as hematite over time.

Warming of early Mars on $10^4-10^7$~year timescales under \ce{O2}-rich atmospheric conditions is not currently predicted in state-of-the-art climate models \cite{methods}, but if it did occur, it could also have led to the formation of diagenetic hematite and manganese in Hesperian sediments. In any case, shorter periods of warming (up to a few years) would still have occurred episodically when the atmosphere was \ce{O2}-rich, via the mobilization of surface ice deposits by redox-neutral bolide impacts \cite{Segura2002,palumbo2018impact}. This would have provided additional routes for the formation, transport and deposition of highly oxidizing minerals. Manganese takes several years to be oxidized by \ce{O2} at moderate ($\sim 280-300$~K) temperatures \cite{Lanza2016}; at higher temperatures, however, the rate increases. Our thermal and kinetic chemical modeling indicates that a few relatively short episodes of elevated surface temperatures in an oxidizing atmosphere with $\sim 10^{-2}$~bar \ce{O2} are sufficient to cause the oxidation of manganese, particularly if the aqueous chemistry occurs under conditions where the water-to-rock ratio is low and the pH is relatively high (Fig.~\ref{fig:Mn_oxidation}). Short intervals of this kind provide a plausible explanation for the Mn-rich fracture-filling materials in the Kimberley Formation at Gale Crater \cite{Lanza2016}.

\subsection*{Model Predictions and Wider Implications}

An episodically fluctuating scenario for Mars' chemical and climate evolution {appears} consistent with several key features of the planet's geological record. It also predicts formation scenarios for various aqueous minerals that can be tested further by NASA's \emph{Perseverance} rover, by the upcoming European and Chinese rover missions and, eventually, by investigation of returned samples on Earth. {First, intermittent warming and redox fluctuations imply multiple erosion episodes of fluvial features like valley networks, rather than formation during a single long warm interval. Second, if TSR was an important early sulfate removal process, reduced sulfur minerals with a petrology consistent with formation at moderate to high temperatures should be present in some Noachian crustal samples. Third, {any} model in which {long-term warming requires reducing atmospheric conditions} predicts specific formation pathways for {hematites and} manganese oxide, which may also be testable via detailed in-situ analysis of samples. Investigation of the {oxidation state} and MIF of sulfur species in terrains of various ages would allow a direct test of the prediction that \ce{O2}-rich atmospheric conditions occurred intermittently in the Hesperian and Amazonian. Finally, because high levels of \ce{CO2} and H input are required in our stochastic model to produce even episodically warm conditions, continued investigation of additional atmospheric warming effects by minor species and aerosols and refined constraints on the change in solar flux over geological time will be valuable.

In the present-day Solar System, the only planet with an oxygen-rich atmosphere is Earth, which has led to suggestions that oxygen could serve as a biomarker gas in the hunt for life on exoplanets. However, our model predicts long-lived (up to $>$100~My), relatively oxygen-rich atmospheres for Mars in the middle period of its history without requiring the presence of life, indicating that \ce{O2} detection alone can be a `false positive' for life in some circumstances. Because prebiotic chemistry does not occur in highly oxidizing environments, this work places constraints on the time periods and locations in which life could have originated and persisted on early Mars (i.e. either during the relatively short warm reducing intervals, or in the subsurface).  This proposed scenario places Mars intermediate to other bodies in the solar system. For example, Venus has undergone extensive water loss to space and likely also oxidation of its surface and mantle, based on its enhanced atmospheric D/H ratio and the limited surface geochemistry constraints that exist \cite{Fegley1997,Wordsworth2016a}. At the other end of the spectrum, Saturn's moon Titan has a strongly reducing atmosphere rich in \ce{CH4} and \ce{H2} \cite{Niemann2005}, despite its low mass. The dynamic history of martian environments proposed here suggests opportunities for the emergence of life during warm, wet intervals when reducing conditions would have favored prebiotic chemistry, but also challenges for the persistence of surface life in the face of frequent and, through time, lengthening intervals of mainly cold and dry oxidizing environments.

\newpage
\section{Methods}

\subsection{Stochastic atmospheric evolution model}

Our atmospheric evolution model is designed to capture the key processes governing atmospheric compositional changes on Mars throughout its geologic history. The model includes terms for supply of reducing species from the subsurface and interior, atmospheric escape and weathering. The key prognostic variable for the model is the oxidizing power of the atmosphere. This is expressed in terms of a single quantity $N$, which we define as the global number of moles of electron acceptors or total oxidizing power present, as in \cite{wordsworth2018redox}. The oxidizing power of a given element is defined under typical planetary conditions, such that O is defined as accepting 2 electrons and H as providing 1, leading to an $N$ contribution of +4 for \ce{O2}, --2 for \ce{H2}, and 0 for \ce{H2O}. We assume $N=0$ at the beginning of each simulation. We work with $N$ in moles, {although sometimes it is also convenient to express it in terms} of the number of electrons accepted by the oxygen in 1~m global equivalent layer (GEL) of \ce{H2O} ice on the martian surface ($1.5\times10^{19}$~mol). The model solves an ordinary differential equation for the evolution of $N$ with time
\begin{equation}
\frac{dN}{dt} = -\left. \frac{dN}{dt} \right|_s +  \left. \frac{dN}{dt} \right|_e - \left. \frac{dN}{dt} \right|_w \label{eq:master_eq}
\end{equation}
where the three terms on the right-hand side represent tendencies due to supply of reducing power from the planet's interior and/or bolide impacts, oxidation via hydrogen loss to space and loss of oxidizing power to the surface via weathering, respectively. 
 
 \subsubsection*{Supply of reducing gases to the atmosphere}
 
 All the main processes that supplied volatiles to the atmosphere of early Mars were highly episodic in nature. To capture this episodicity in a general way, we developed a stochastic representation of the injection of reducing gases to the atmosphere, with the value of ${dN}/{dt} |_s$ sampled at every timestep from a power law distribution. A power law distribution is a physically realistic choice, because similar statistics are observed in a wide range of geophysical settings, including specifically both bolide impacts and volcanism on planetary bodies [e.g., \cite{tremaine1993statistical,werner2011redefinition,mason2004size,cannavo2016possible}]. Keeping the representation of the reducing gas supply general also has the advantage of allowing systematic study of the behaviour of the model when the governing parameters are varied. {To avoid unnecessary complexity in the model, we assume zero autocorrelation in the forcing.}
 
The power law distribution over some interval $x \in [x_0,x_1]$ is defined as 
\begin{equation}
p(x) = Cx^\gamma,\tag{\ref{eq:probeqn}}
\end{equation}
with the normalization requirement
\begin{equation}
\int_{x_0}^{x_1} p(x) dx = 1 \tag{\ref{eq:probnorm}}
\end{equation}
allowing us to define 
\begin{equation}
C = \frac{\gamma+1}{ x_1^{\gamma+1}  -  x_0^{\gamma+1} }.
\end{equation}
 The expectation value of this distribution, $\mu$, is
\begin{equation}
\mu \equiv E[x] =  \frac{\gamma+1}{\gamma+2} \frac{x_1^{\gamma+2}-x_0^{\gamma+2}}{x_1^{\gamma+1}-x_0^{\gamma+1}} 
\label{eq:PLmean}
\end{equation}
and the variance $\sigma^2$ is 
\begin{equation}
\sigma^2 \equiv E[(x-\mu)^2] =  \frac{\gamma+1}{ x_1^{\gamma+1}  -  x_0^{\gamma+1} }  \left( \frac{x_1^{\gamma+3}-x_0^{\gamma+3}}{\gamma+3} - 2\mu \frac{x_1^{\gamma+2}-x_0^{\gamma+2}}{\gamma+2} + \mu^2 \frac{x_1^{\gamma+1}-x_0^{\gamma+1}}{\gamma+1}  \right).\label{eq:PLvar}
\end{equation}
Both of these quantities remain finite-valued as long as $x_0$ and $x_1$ are finite and $\gamma$ is not exactly equal to -1, -2 or -3.  

In our model, $x$ is the supply rate of reducing power to the atmosphere in mol/s. Hence $\mu$ is the mean supply rate, which we allow to vary slowly over geological time to account for the secular decline in all reducing gas supply processes (see below). $x_1$ is taken to be the largest possible flux of reducing power in a single timestep, which we also estimate below. Given $\mu$ and $x_1$, $x_0$ is then calculated from equation \eqref{eq:PLmean} via a root-finding algorithm. 

The distribution is sampled in the model by first selecting a random number from the uniform distribution on the domain $x \in [0,1]$. Then the sampled value from the power law distribution is 
\begin{equation}
x_{s} = [(x_1^{\gamma+1} - x_0^{\gamma+1})x + x_0^{\gamma+1}]^{1/(\gamma+1)}
\end{equation}
The supply tendency is then simply
\begin{equation}
\left. \frac{dN}{dt} \right|_s = x_s.
\end{equation}
or
\begin{equation}
\left. \Delta N \right|_s = x_s \Delta t.
\end{equation}
in the discretized model with timestep $\Delta t$. 

Along with their short-term variability, all the major mechanisms for supply of reducing gases to the martian atmosphere also underwent secular decay with time, because the interior and crust cooled from the early Noachian onwards and the bolide impactor flux declined after the main period of planetary accretion. We represent this decline in our model via a simple exponential decay law
\begin{equation}
\mu(t) = \mu_0\mbox{e}^{-t/\tau}.\label{eq:sec_decay_law}
\end{equation}
Here $\mu_0$ and $\tau$ are model parameters that we allow to vary within physically constrained ranges. The absolute time mean is then defined as 
\begin{equation}
\overline \mu = \frac 1{t_n} \int_0^{t_n}\mu(t) dt = \mu_0  (1-\mbox{e}^{-t_n/\tau})(\tau/t_n) \label{eq:time_mean}
\end{equation}
and  the total integrated gas injection value as $N_s = \overline \mu t_n$, where $t_n = 4.5$~Gy is taken as the current age of Mars. 

To build a comprehensive picture of the range of plausible climate scenarios on early Mars, we tested the effects of varying $\overline \mu$ and $x_1$ and other parameters over a range of values (Table~S1). For Figure~\ref{fig:pspace_span}, an ensemble of 256 simulations with identical initial and boundary conditions was created for each $\overline \mu$ and $\beta  \equiv x_1 \slash \overline \mu$ value studied. The colored lines in the plot correspond to the ensemble mean values for each case. We also tested the effect of varying $\gamma$ and $\tau$ (Fig.~S4). The range for $\tau$ was determined based on timescales in volcanic and impactor flux models [e.g., \cite{Grott2011,werner2011redefinition}]. The range for $\gamma$ was determined based on values in standard impactor flux models \cite{tremaine1993statistical,Sleep2001,werner2011redefinition} and scalings for volcanism on present-day Earth \cite{mason2004size,cannavo2016possible,papale2018global}.  For bolide impacts, the data suggests $\gamma$ is between $-1.5$ and $-2$ in the solar system \cite{zahnle1992impact}. For volcanism, \cite{papale2018global} argue for a power law distribution with $\gamma = -1.95$, \cite{cannavo2016possible} argue for $\gamma = -2.16$, while comparison with the distribution proposed in \cite{mason2004size} with equivalent power law distributions suggests $\gamma \approx -3.1$. 
Here we use $\gamma = -1.95$ as our nominal value but also study sensitivity to both lower and higher values (Fig.~S3). For crustal alteration and clathrate release, $\gamma$ is difficult to constrain, but a similar range is expected if release events are triggered by bolide impacts or volcanism. 

\subsection*{Constraints on reducing gas release}

We produced the generalized constraints on reducing gas release in Figure~2 and accompanying discussion in the main text as follows. 

\textbf{Volcanism:} Grott et al. \cite{Grott2011} estimated a total volcanic outgassing of {17.5-61}~m~GEL of \ce{H2O} throughout martian history assuming a mantle \ce{H2O} concentration of 100~ppm, depending on whether a global melt channel or localized mantle plume melting is assumed. Later, \cite{McCubbin2012} estimated a martian mantle \ce{H2O} content range from 73 to 290~ppmw based on a geochemical analysis of apatites in shergottite meteorites. Finally, oxygen fugacity ($f_{\ce{O2}}$) values in martian meteorites range from around the iron-w\"ustite (IW) buffer to the  quartz-fayalite-magnetite (QFM) buffer, which translates to outgassed {molar} ratios of \ce{H2} to \ce{H2O} from around 0.65 to 0.025 \cite{Wadhwa2001,Herd2003,Ramirez2014,wordsworth2018redox}. Taken together, these limits imply a range of 12.25 to 176.9~m total \ce{H2O} outgassed and an average \ce{H2} outgassing rate over 4.5~Gy from $33.2$ to $1.2\times10^{4}$~mol {H}/s. These latter two values correspond to the lower and upper limits for  $\overline \mu$ for volcanism given in Figure~2.  For comparison, the hydrogen outgassing rate required for steady-state warming via an atmospheric \ce{H2} molar concentration of 5\%, assuming escape at the diffusion limit, is $2.2\times10^{6}$~mol H/s. Even assuming all volcanism occurred over the 700~My period from 4.2 to 3.5~Gya, this value is over an order of magnitude higher than the upper limit given above.

Constraining the upper limit for $\beta$ from volcanism is challenging given the uncertainties in martian geodynamics and eruptive processes, but an order-of-magnitude estimate can be gained by considering the Tharsis Rise, which was formed during the Noachian. The total volume of Tharsis has been estimated as around 2000~m GEL of basalt \cite{Phillips2001}. Given a basalt density of 3000~kg/m$^3$ and upper limit melt \ce{H2O} specific concentration of 2870~ppmw \cite{McCubbin2012}, this implies release of $1.4\times10^{20}$~mol of \ce{H2O}, or $3.6\times10^{18}$~mol to $9.4\times10^{19}$~mol of \ce{H2} for $f_{\ce{O2}}$ at IW to QFM. Achieving $\beta = 10^2$ to $10^3$ requires a maximum of $1.2\times10^{6}$ to $1.2\times10^{7}$~mol~H release in a 100,000~y interval, which translates to a few to tens of percent of the volume of Tharsis created in this time period. This is challenging, but might not be impossible during a period of particularly intense volcanic activity.

\textbf{Bolide impacts:} To derive the upper limit on H input from bolide impacts in Fig.~2, we used the methodology of \cite{Haberle2019}. We used the full set of impact crater radii listed in their Supplementary Information and calculate the impactor radius as $r = 0.022 D_c^{1.076}$~m, where $D_c$ is the crater radius in metres, based on their eqns. S1 and S3. We assume that the impactors were composed of H-chondrite material with a metallic iron mass fraction of 27\% and use an iron--water reaction to form \ce{H2} as \ce{Fe + H2O \to FeO + H2}. The contribution of shocked crustal material to H release is neglected in this calculation. The calculation gives an average H input over 4.5~Gy from impacts of $2.47\times10^4$~mol H/s. Regarding episodicity, a sense of the numbers can be gained by considering Antoniadi Crater, which is a `typical' moderate-sized Noachian crater with diameter 381~km, yielding an estimated bolide impactor radius of 34.8~km. Using the above assumptions on H release gives $8.9\times10^{17}$~mol~H, equivalent to $2.8\times10^5$~mol~H/s in a 100,000 year model timestep or $\beta \approx 11.4$. In contrast, for the biggest impactor craters such as Argyre (1315~km) and Hellas (2070~km), which are dated to the early to mid Noachian \cite{Haberle2019}, we find $\beta \approx 620$ and $2680$, respectively. Hence we conclude that direct bolide impact H creation can explain the formation of fluvial features and crater degradation in the early to mid Noachian, but probably not in the late Noachian / early Hesperian.

\textbf{Crustal alteration:} Processes such as serpentinization and radiolytic \ce{H2} production likely caused extensive episodic outgassing of \ce{H2} and related species like \ce{CH4}  \cite{Chassefiere2016,Chassefiere2013,Batalha2015,tarnas2018radiolytic}. Globally, it has previously been estimated that up to 300-1000 m GEL of water could have been consumed by serpentinization of Mars' Noachian crust \cite{Chassefiere2013}. 
Assuming that the ratio of \ce{H2O} consumed to \ce{H2} emitted is around 7, this implies 42.9 to 142.9~m~\ce{H2O}~equiv. of \ce{H2}, or $4.5\times10^3$ to $1.48\times10^4$~mol~H/s averaged over 4.5~Gy. Emission rates peaking at up to $5\times10^{14}$ to $5\times10^{16}$~molecules/m$^2$/s via serpentinization of mafic and ultramafic minerals could have occurred locally \cite{Wordsworth2017}, based on observations of emissions from terrestrial ophiolites \cite{Etiope2013}. Formation of \ce{CH4} clathrates could lead to instantaneous reduced gas release rates that are even higher than this, depending on the efficiency of methane formation during serpentinization and other factors \cite{Wordsworth2017,Chassefiere2016,Lasue2015}.  Finally, positive crustal feedbacks on H release may occur once warming begins due to authigenic magnetite formation and associated hydrogen release in surface aqueous environments \cite{tosca2018magnetite}. The effect of this process is hard to quantify, but it would likely strongly increase the peak hydrogen flux following short-term thermal perturbations by bolide impacts.

\subsubsection*{Atmospheric escape}

The second term in equation~\eqref{eq:master_eq} describes net change in atmospheric oxidizing power via escape of hydrogen and (to a lesser extent) oxygen to space. We include a number of processes to determine the variation of this term with time.  

\textbf{Molecular and atomic hydrogen escape:} We assume that when \ce{H2} is present in the atmosphere ($N<0$ in our model), its loss is diffusion-limited following the formula for a general diffusing species $i$
\begin{equation}
\Phi_{diff-lim} = b_{i}x_i\left(\overline H^{-1} - H_i^{-1}\right) .\label{eq:difflimesc}
\end{equation}
Here $\overline H$ is the scale height of the background atmosphere, $H_i$ is the scale height of \ce{H2}, $x_i$ is the \ce{H2} molar concentration and $b_i$ is the binary diffusion coefficient for \ce{H2} and \ce{CO2}, which we take from \cite{Chamberlain1987}. 

When \ce{H2} is present in the atmosphere, equation \eqref{eq:difflimesc} determines the oxidation rate. However, oxidation still occurs even in the absence of atmospheric \ce{H2} if photodissociation of \ce{H2O} can lead to non-stoichiometric hydrogen escape.  On present-day Mars, convective lofting during global dust storms is a key contributor to non-stoichiometric H loss today \cite{heavens2018hydrogen}.  Propagation of volatile species to the upper atmosphere of any planet is a strong function of orbital obliquity, the dust cycle, surface ice distribution and surface pressure, so during periods in the past where these parameters varied,  transport of \ce{H2O} to the high atmosphere could have been increased by many orders of magnitude. This is consistent with Mars' high degree of water loss based on the observed deuterium to hydrogen (D/H) ratio through time.

Here we incorporate an average transport rate of \ce{H2O} to the high atmosphere during cold periods that permits the model to match D/H values through time. Mars' D/H ratio today is about $6$ times standard mean ocean water (SMOW) 
\cite{donahue1995evolution,villanueva2015strong,heard2020prob}. The D/H value around 3.5~Gya was $3\pm0.2$ times SMOW, 
based on analysis of sedimentary rocks at Gale crater \cite{mahaffy2015imprint}, while Mars' formation D/H value was likely 1.2 to 1.6 times SMOW \cite{lunine2003origin}. 
The \ce{H2O} inventory in moles at time $t$ from initial delivery is derived from the D/H ratio as
\begin{equation}
N_{\ce{H2O}}(t) = N_{\ce{H2O}}(t_n)\left( \frac{r_{D-H}(t_n)}{r_{D-H}(t)} \right)^{1/(1-f)} 
\end{equation}
\cite{donahue1995evolution,kurokawa2014evolution}, where $t_n$ is the time today (4.5~Gy), $r_{D-H}(t)$ is the D/H ratio at time $t$, and $f$ is the fractionation parameter. 
Here we take $f=0.016$, 
which leads to an estimate of {an initial 130-170~m~GEL \ce{H2O} inventory on Mars} \cite{yung1988hdo,krasnopolsky1998detection,mahaffy2015imprint,heavens2018hydrogen}. We assume that the time-dependence of $N_{\ce{H2O}}(t)$ is 
\begin{equation}
N_{\ce{H2O}}(t) =  \frac 12 \alpha L_{\ce{H},n} \tau_{\ce{H2O}} (\mbox{e}^{-(t - t_n)/\tau_{\ce{H2O}}} - 1) + N_{\ce{H2O}}(t_n).
\end{equation}
 
Here $N_{\ce{H2O}}(t_n) $ is the estimated present-day surface \ce{H2O} inventory ($2.5\times10^{20}$~mol or 34~m~GEL~\ce{H2O}) \cite{Carr2015} 
and $L_{\ce{H},n} = 1.7\times 10^{3}$~mol/s (3.6~m~\ce{H2O}~equiv./Gy) 
is the present-day loss rate of H to space. $\tau_{\ce{H2O}} $ and $\alpha$ are free parameters that we determined to be $1.11$~Gy and $0.48$, respectively, {using an L2 norm error minimization procedure} vs. observational D/H data. This parametrization correctly reproduces the initial, 3.5~Gya and present-day martian D/H ratios, {within error bounds} (Fig.~S1).

The change in overall oxidizing power vs. time from the associated H loss in our model is then
\begin{equation}
\left. \frac{d N}{dt} \right|_{e,\ce{H2O}} = -2  \frac{d N_{\ce{H2O}}}{dt} = +  \alpha L_{\ce{H},n}  \mbox{e}^{-(t - t_n)/\tau_{\ce{H2O}}} .
\label{eq:waterloss}
\end{equation}
For simplicity,  we neglect the effects of deep crustal interaction with \ce{H2O} on D/H variations here [e.g., \cite{Chassefiere2013}]. The importance of this process remains unclear, although if early Mars was mainly cold, the subsurface cryosphere would have placed strong constraints on the rate of isotopic exchange between deep groundwater and the surface \ce{H2O} inventory.

\textbf{Non-thermal O escape:} We also include non-thermal escape of O to space in our model. The escape flux of all O species observed from Mars today by MAVEN is around $6\times10^{25}$~atom/s \cite{jakosky2018loss,brain2015spatial,lillis2015characterizing}. 
Past escape rates of O (from both \ce{H2O} and \ce{CO2}) were likely larger via a combination of enhanced dissociative recombination, ion escape, and ion sputtering \cite{jakosky1994mars,Chassefiere2004,Lammer2013,jakosky2018loss}.  {Here we include the effect of non-thermal loss on redox balance in our model by assuming the dominant loss channel is dissociative recombination and using the formula from \cite{lillis2017photochemical} with power law coefficient of $1.7$ \cite{heard2020prob}. Before 3.5~Gya, we fix the maximum loss rate at $1\times10^{27}$, based on \cite{jakosky2018loss}. We also calculated the drag of oxygen with escaping hydrogen when the extreme ultraviolet (XUV)-driven escape flux was high, but found that it was small from the early Noachian onwards and therefore neglected it. 

Finally, all escape fluxes $\Phi$ are converted to a global rate of change in mol/s as
\begin{equation}
\left. \frac{dN}{dt} \right|_e = 4\pi r^2  \Phi / N_A
\end{equation}
where $r$ is the radius of Mars and $N_A$ is Avogadro's constant. {Because the input of reducing gas in the model is stochastic, the percentage of time the atmosphere is oxic is a function of both $\overline \mu$ and $\beta$. Figure~S2 shows the effect of varying $\beta$ on the model output, while Fig.~S3 shows the total percentage of time the atmosphere is oxic for Noachian and for Hesperian + Amazonian time periods as a function of $\overline \mu$ and $\beta$.}

\subsubsection*{Evolution of \ce{CO2} with time}

The evolution of \ce{CO2} in Mars' atmosphere is taken into account using constraints from spacecraft atmospheric escape observations, and models of impact ejection, carbon isotopic evolution and crustal sequestration \cite{jakosky2019co2}. 
We use the following empirical function for the variation of $p_{\ce{CO2}}$ with time:
\begin{equation}
p_{\ce{CO2}} = A(1 - \tanh [a(t-0.8~\mbox{Gy})]).  \label{eq:pCO2}
\end{equation}
Here $t$ is in Gy. In our standard scenario, $a=1.5$~Gy$^{-1}$ and {$A = 2.1$~bar, which leads to 0.46~bar 3.0 Gya, 1.5~bar 3.5 Gya and 2.99~bar 4.0~Gya. This is consistent with the review of \cite{jakosky2019co2}, which estimates $\sim 1.5$ to $>3$~bar of \ce{CO2} has been lost from the martian atmosphere since 4.3~Gya, primarily due to a combination of impact erosion, stripping to space and deep crustal carbonate formation. It is also consistent with paleo-atmosphere modeling constraints based on observed crater size distributions \cite{Kite2014}, which yield an upper limit for atmospheric pressure of 1-2~bar 3.6~Gya. Finally, it is within the range of upper limits for Noachian $p_{\ce{CO2}}$ from C isotope fractionation modeling, as long as the bulk of subsurface carbonate formation is not driven by evaporative processes \cite{Hu2015}. Sensitivity tests assuming lower and higher values of $A$ are presented in Figure~S4.

\subsubsection*{Removal of atmospheric \ce{O2} via surface processes}

Because the martian crust and mantle contain large quantities of reduced species (particularly iron in the \ce{Fe^{2+}} oxidation state), interaction of any liberated \ce{O2} in the martian atmosphere with the surface is a potentially important sink of oxidizing power (decrease in $N$ towards zero). Sinks of \ce{O2} due to aqueous processes during brief {($\sim$a few years)} warming events are likely to be small (see Supporting Information). When Mars is in a cold climate state, the rate of iron oxidation is orders of magnitude lower than during warm periods, but probably still non-zero \cite{burns1993rates}. Processes such as oxidation during local transient melting events at the surface \cite{haberle2001possibility,mcewen2014recurring}, alteration of hydrothermal fluids in intermittent contact with the atmosphere and possibly dry photo-oxidation \cite{huguenin1973photostimulated} may all have contributed to slow removal of \ce{O2} and related oxidants from the atmosphere. We included a small constant flux of oxidizing power out of the atmosphere during periods where $N>0$, although we found it had little effect on the duration of warm periods in the model.} The default value {we use in the simulations} is $ dN/dt|_w =  1.6\times10^{3}$~mol/s (equivalent to the H in 15~m~GEL of \ce{H2O} over 4.5~Gya), although a wider range of values was studied in sensitivity tests (see Table~S1).

\subsubsection*{Climate parametrization}

We use the PCM\_LBL line-by-line radiative-convective code \cite{Wordsworth2017} to calculate equilibrium surface temperature as a function of reducing gas concentration in the model. For simplicity, we assume that \ce{CO2} and \ce{H2} are the primary warming agents, although other gases such as \ce{CH4} and longer chain hydrocarbons may also have made important contributions that we neglect here. Surface temperatures are calculated as a function of solar flux, total surface pressure and hydrogen molar concentration. All other parameters were set as in \cite{Wordsworth2017}. Simulations were run for 500 steps, which was sufficient to achieve equilibrium in all cases. The resulting array was then used to produce surface temperature values during operation of the stochastic climate model via linear interpolation. We tested the effect of using \ce{CO2-H2} collision-induced absorption values from both \cite{Wordsworth2017} and \cite{turbet2020measurements}. The latter values are slightly lower, so to remain conservative in our predictions of surface warming we used them for the results reported here. {Our results closely match those of \cite{turbet2020measurements}, which found surface temperature $T_{s} = 273$~K at  2~bar \ce{CO2} and 6\% \ce{H2} using the correlated-$k$ model described in \cite{wordsworth2010}. Our calculated surface temperatures at 0\% \ce{H2} are warmer by a few Kelvins than previous correlated-$k$ calculations because our line-by-line model produces a slightly cooler upper atmosphere [an effect seen already in other intercomparisons, e.g. \cite{ding2019}].} Solar luminosity through time was calculated using the formula from \cite{gough1981}. We also tested the effect of assuming constant solar luminosity at the present-day value in some runs as an upper limit implementation of the `massive young Sun hypothesis' \cite{minton2007} (Fig.~S4). 

\subsection{Short-term regolith heating from bolide impacts}

The initial effect of large impacts is to cause a short intense pulse of regional to global surface heating by mobilizing ice and rock as liquid and vapor ejecta that is subsequently deposited on the surface (see Supporting Information). To calculate the sub-surface warming due to emplacement of a hot surface layer, we used a regolith thermal evolution model, with the upper boundary condition either Stefan-Boltzmann cooling or surface temperatures from the results of 3D global climate modeling in the aftermath of a giant impact \cite{steakley2019testing}. The regolith thermal evolution model solves the heat equation
\begin{equation}
c_h\rho\frac{\partial T}{\partial t} =  \frac{\partial}{\partial z}\left(\kappa_r \frac{\partial}{\partial z} T\right)  \label{eq:heat}
\end{equation}
where $T$ is temperature, $t$ time, $z$ depth, $\kappa_r$ the regolith thermal conductivity, $c_h$ heat capacity (840~J/K/kg here) and $\rho$ density (3000~kg/m$^3$ here).  Each simulation is initialized with a hot upper layer overlying a colder layer representing pre-impact regolith in geothermal equilibrium. The hot upper layer is between 1 and 500 m deep and is assigned a uniform temperature of 800 to 1600~K. The cold lower layer is initialized at 220~K at its shallowest level, with higher initial values with depth according to the geothermal gradient. We use a geothermal gradient of 32.5~K/km based on an assumed Noachian geothermal heat flux of $F_{geo} = 65$~mW/m$^2$ and a thermal conductivity of 2~W/m/K \cite{Clifford1993}.  Latent heat was incorporated via the apparent heat capacity method \cite{hu1996mathematical}, using a saturated regolith with 20 \% porosity. The bottom boundary condition is set by $F_{geo}$. For the upper boundary condition, Stefan-Boltzmann cooling was used to give a conservative upper limit on the cooling rate.

\subsection{Sulfur chemistry modeling}

We produced the pH--$f_{\ce{O2}}$ equilibrium plot for sulfur minerals in Fig.~3A using \emph{The Geochemist's Workbench} software package. The magnetite-hematite and quartz-fayalite-magnetite mineral redox buffers were calculated using the same approach as in \cite{wordsworth2018redox}. 
The thermochemical sulfate reduction calculation for Fig.~3B was performed by coupling the one-dimensional thermal diffusion model discussed above to a chemical solver. We used existing experimental data on thermochemical sulfate reduction obtained at \ce{H2} partial pressures of 4-16 bar \cite{truche2009experimental} to parameterize this process. We express the removal rate of sulfate as 
\begin{equation}
k(T) = k_0 \mbox{e}^{-\frac{E_a}{R}(1/T - 1/T_{ref})} \quad \mbox{s}^{-1} \label{eq:TSR}
\end{equation}
where $k_0 = 7 \times 10^{-8}$~s$^{-1}$, 
$R$ is the ideal gas constant, $T_{ref} = 280$~$^\circ$C and $E_a = 131$~kJ/mol. The dependence of the reaction rate on \ce{H2} pressure is relatively weak \cite{truche2009experimental} and was therefore neglected here.  

 The decrease in the local concentration of sulfate $C_{\ce{SO4}}$ as a function of time is expressed as 
\begin{equation}
\frac{d C_{\ce{SO4}} }{dt} = -k(T)C_{\ce{SO4}}\label{eq:S_react}
\end{equation}
The fraction of sulfate that becomes reduced as a function of depth $f_{S red}$ is then calculated by integrating equation \eqref{eq:S_react} to get
\begin{equation}
f_{S red}(z,t_f) \equiv C_{\ce{SO4},t_f} / C_{\ce{SO4},0} =  \mbox{ exp}\left[-\int_0^{t_f} k[T(z,t)] dt \right].
\end{equation}
We use the subsurface thermal evolution data described in the last section to obtain temperature and hence $k$ as a function of depth and time at the end of the simulation $t_f$. As discussed above, $t_f$ is pre-determined in the thermal simulations as a function of impactor size. The effective depth to which sulfate is reduced is then calculated for a given hot layer depth by integrating $f_{S red}$ in depth $z$
\begin{equation}
d_{S red} =  \int_0^{z_m} f_{S red}(z,t_f) dz. 
\end{equation}
where $z_m$ is the lowest layer depth in the model. We neglected the dissolution of sulfate minerals in the calculation on the basis that it should proceed much faster than the thermochemical reduction itself. 

Aqueous solutions of sulfate in open-water systems near the surface can be reduced by dissolved \ce{H2} or other reducing gas species from the atmosphere itself. For sulfates that percolate deeper into the regolith, the transport rate of gas species from the atmosphere will eventually become limited, but TSR can also be driven by ferrous iron from crustal minerals \cite{Lorand2020}. As is clear from Fig.~3A, pyrite and pyrrhotite are the stable forms of sulfur when the oxygen fugacity of the surrounding fluid is at the magnetite-hematite buffer value or lower. Most early martian regolith would have been much more reducing, probably at oxygen fugacity between QFM and IW+1. \ce{H2_{(aq)}} reduces minerals such as hematite slowly compared to other reductants, which suggests that the TSR reaction is likely to proceed faster in the deeper regolith than \eqref{eq:TSR} suggests, but further experimental constraints on TSR in this regime would be useful. Finally, we do not consider sulfate reduction in the sub-surface over long time periods at more moderate temperatures in our TSR calculation, although this process may also have been important, particularly in the early to mid-Noachian.

\subsection{Manganese oxidation modeling}

The pH--$f_{\ce{O2}}$ equilibrium plot for manganese minerals in Fig.~4A was produced using \emph{The Geochemist's Workbench}. 
For Fig.~4B, we modeled the kinetics of manganese oxide formation as follows.  In the presence of dissolved oxygen (\ce{O2}), manganese in aqueous solution oxidizes slowly via reactions such as  
\begin{equation}
\ce{4 Mn^{2+}_{(aq)} + O2_{(aq)}  + 6H2O \to  }  \ce{ 4 Mn OOH_{(s)} + 8H^+}
\end{equation}
and
\begin{equation}
\ce{3 Mn^{2+}_{(aq)} + \frac 12 O2_{(aq)} + 3H2O \to Mn_3O_4_{(s)} + 6H^+}.
\end{equation}
The oxidized forms of Mn produced in these reactions are relatively insoluble and subsequently fall out of the solution to form minerals such as hausmannite. We modeled the oxidation of manganese during oxidizing atmospheric intervals in the short-term aftermath of a bolide impact using reaction rates from the experimental literature \cite{hem1981rates,morgan2005kinetics}. Surface temperature values vs. time for impactors of radius 30~km and atmospheric \ce{CO2} concentrations from 0.15 to 1~bar were taken from 3D global climate model output, using simulations performed as described in \cite{steakley2019testing}.

We assume that oxidation of aqueous Mn is the rate-limiting step for Mn oxide formation, and model it via the representative equation
\begin{equation}
\frac{d  \ce{[Mn^{2+}_{(aq)}]}  }{dt}  = - k_{app}(T) K_H(T) p_{\ce{O2}} \ce{[Mn^{2+}_{(aq)}]}= - \frac{ \ce{[Mn^{2+}_{(aq)}]}}{\tau_{Mn Ox}} \label{eq:Mn_react}
\end{equation}
where \ce{[Mn^{2+}_{(aq)}]} is dissolved manganese concentration, $p_{\ce{O2}}$ is the atmospheric \ce{O2} partial pressure, $K_H$ is the Henry's law coefficient for \ce{O2} and $k_{app}$ is an effective reaction rate. Here we take $K_H = K^o_H \mbox{e}^{\alpha(T^{-1} - T_0^{-1})}$, where $K^o_H = 1.3\times10^{-8}$~mol/kg/Pa, $\alpha = 1700$~K and $T_0 = 298.15$~K \cite{linstrom2001nist}. We assume a room-temperature value for $k_{app}$ of $1\times 10^{-4}$~kg/mol/s, based on experiments carried out at a pH of 8 and assuming 1~kg/L of \ce{H2O} \cite{morgan2005kinetics}.  
Manganese oxidation proceeds faster at higher pH values, but pH = 8 was judged to be a reasonable value for reactions occurring in the presence of basaltic crust at relatively low water to rock ratios. Manganese oxidation also proceeds faster when self-catalysis can occur. We do not model this process explicitly, but we note that the high temperatures immediately following a bolide impact would allow rapid oxidation of small quantities of Mn, likely driving effective subsequent self-catalysis. The temperature dependence of $k_{app}$ is taken into account by assuming Arrhenius rate dependence and calculating the activation energy from experimental data at a fixed pH of 8 [specifically, we use experiments K13 and K17 from \cite{hem1981rates}]. Using this data, we derive an effective activation energy of $E_a = 182.6$~kJ/mol.  



\newpage

\section*{Corresponding Author}

Correspondence and requests for materials should be addressed to R. Wordsworth:\\ {\emph{rwordsworth@seas.harvard.edu}}.

\section*{Acknowledgments}
RW thanks D. Johnston, K. Loftus, Y. Sekine and K. Zahnle for discussions. We thank N.~Lanza, E.~Kite, D.~Catling and N.~Mangold for their helpful comments during peer review. 
\textbf{Funding:} RW and MB acknowledge funding from NSF CAREER award AST-1847120 and NASA/VPL grant UWSC10439. JAH acknowledges support from the Simons Collaboration on the Origin of Life.
\textbf{Author contributions:} RW, AK and JH conceived the paper. MB performed the 1D regolith heating calculations. KS performed the post-impact 3D climate simulations. RW performed the rest of the modeling and wrote most of the manuscript. 
BE provided ideas and text on surface geochemical evolution, timing, and the role of evaporitic vs. thermochemical processes. JWH provided input and background on planetary volcanism, impactor and fluvial/lacustrine processes, and geological evolution. All authors discussed the results and provided input on multiple draft versions of the manuscript. 
\textbf{Code availability:} The stochastic atmospheric evolution model, along with other scripts to reproduce plots in the paper, is available open-source at \emph{https://github.com/wordsworthgroup/mars\_redox\_2021}. The line-by-line radiative-convective model used to produce the climate parametrization (PCM\_LBL) is available open-source at 
\\\emph{https://github.com/wordsworthgroup/mars\_redox\_2021/tree/main/PCM\_LBL}. 
The regolith thermal evolution model is available open-source at \emph{https://github.com/wordsworthgroup/crustal-heat}. \emph{The Geochemist's Workbench} is proprietary software available at \emph{https://www.gwb.com/}. 
\textbf{Data availability:} The input data to the model used to produce the plots in this paper is contained within the \emph{github} repositories listed above, with the exception of the HITRAN data used by PCM\_LBL and the 1D regolith heating data, which are available at \emph{10.5281/zenodo.4458673}. 

\begin{figure}[h]
	\begin{center}
		{\includegraphics[width=3.75in]{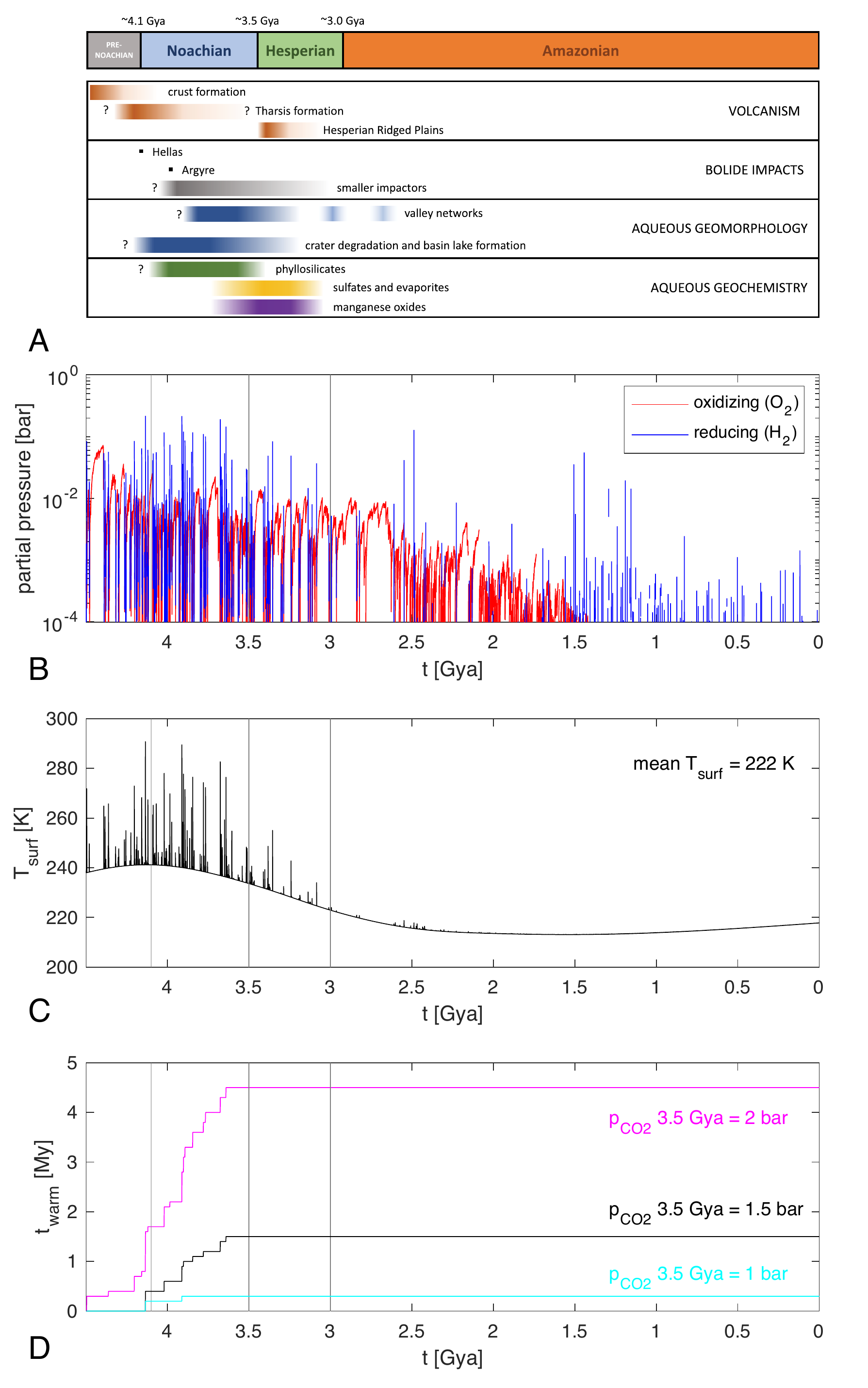}}
	\end{center}
	\caption{\textbf{Observations vs. atmospheric evolution model predictions over Mars' history. } A) Timeline of major events on the martian surface from geologic observations (here Gya means gigayears ago). B) Changes in the net redox state of the atmosphere vs. time due to the competing effects of episodic release of reducing gases, atmospheric escape and surface weathering. C) Corresponding surface temperature evolution and D) cumulative integral of time spent with surface temperature $>273$~K ($t_{warm}$), for three assumed \ce{CO2} evolutionary tracks (nominal value of $p_{\ce{CO2}}$~3.5~Gya~$= 1.5$~bar). In the nominal run, the time-mean surface temperature is 222~K, and the integrated period of warm conditions from the Noachian onwards is a little over 1~My. }
\label{fig:coupled}
\end{figure}

\begin{figure}[h]
	\begin{center}
		{\includegraphics[width=6.5in]{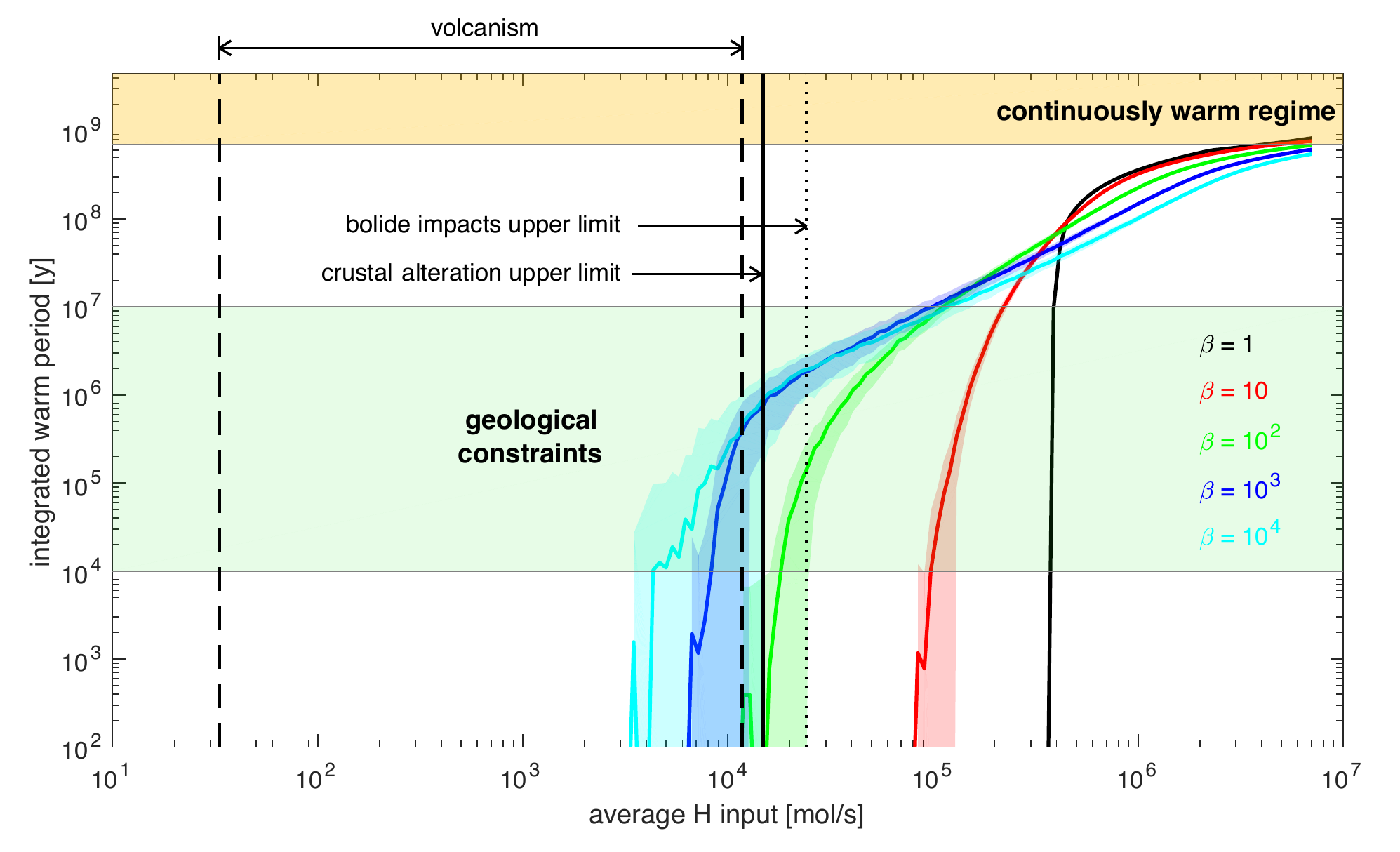}}
	\end{center}
	\caption{
\textbf{Integrated duration of warm intervals as a function of H fluxes into the martian atmosphere.} Colored lines show the integrated warm period from 4.1~Gya onwards (ensemble mean {of 256 runs at each point}) as a function of the average rate of reducing gas (\ce{H2}) input to the atmosphere over the entire 4.5 Gy simulation interval. Colors indicate  the {assumed variability ratio $\beta$} [specifically, the maximum possible input rate in a 0.1 My interval relative to the {time-varying} mean value {$\overline \mu(t)$}]. The central bright lines show mean values, while the lighter shading indicates 1-$\sigma$ deviation from the ensemble mean. Results show that when source variability is high (cyan, blue and green lines), this strongly increases the total warm duration for physically reasonable average H input rates (see black vertical lines). }
\label{fig:pspace_span}
\end{figure}

\begin{figure}[h]
	\begin{center}
		{\includegraphics[width=6.5in]{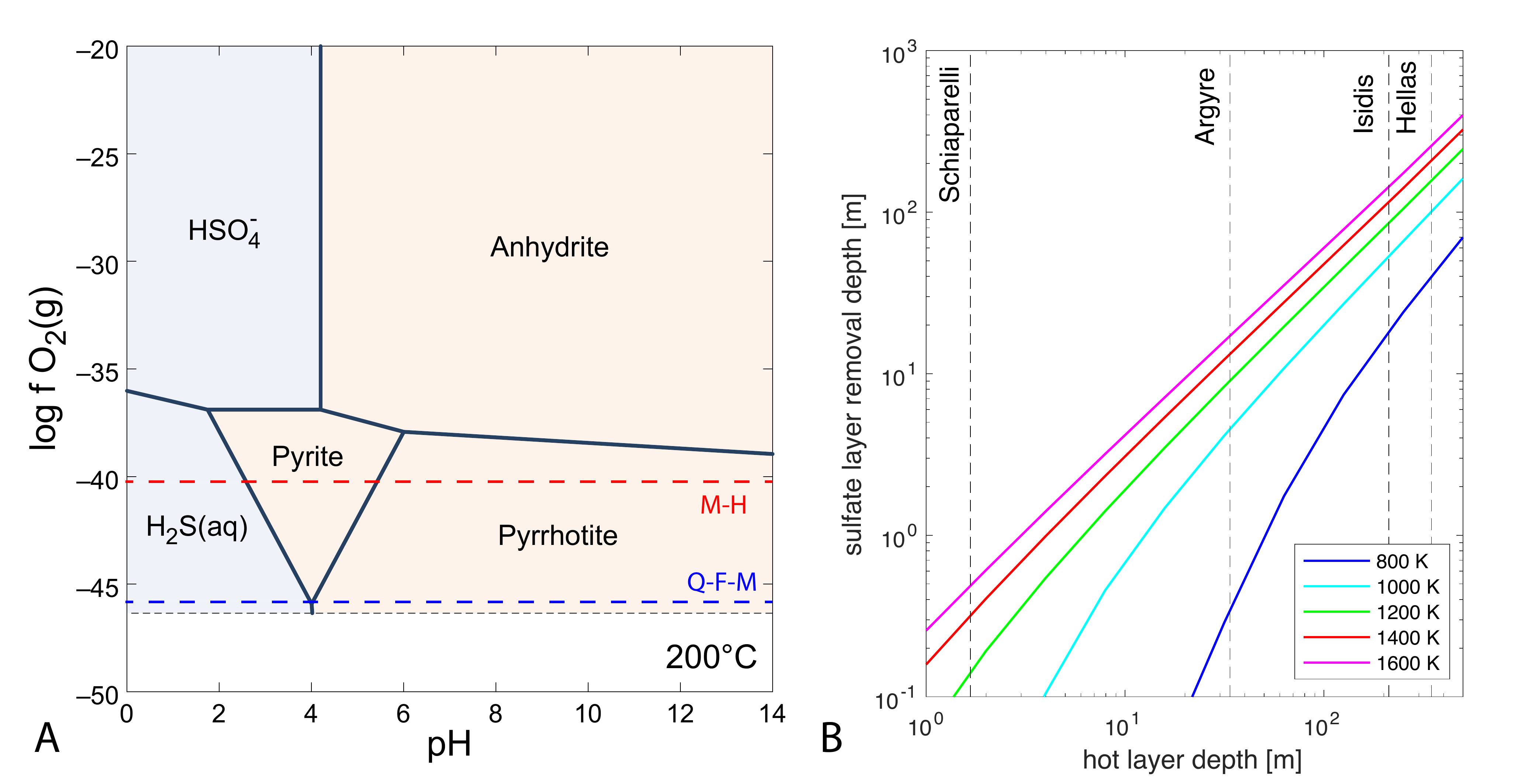}}
	\end{center}
	\caption{\textbf{Thermochemical sulfate reduction on Mars in the Noachian period.} A) Equilibrium aqueous chemistry of sulfur as a function of pH and oxygen fugacity, assuming a temperature of $473$~K (200~$^\circ$C) and \ce{[SO_4^{2-}]} = \ce{[Fe^{2+}]} = \ce{[Ca^{2+}]} = $10^{-3}$. {Locations of the magnetite-hematite and quartz-fayalite-magnetite buffers are shown in red and blue, respectively.} B) Depth of surface layer in which sulfate {can be} reduced to pyrite {or} pyrrhotite following emplacement of a hot surface layer. Results are shown as a function of the hot layer depth and initial temperature. The estimated global mean ejecta depths from \cite{Toon2010} due to a few known impactors are also shown.}
\label{fig:S_thermochem}
\end{figure}

\begin{figure}[h]
	\begin{center}
		{\includegraphics[width=6.5in]{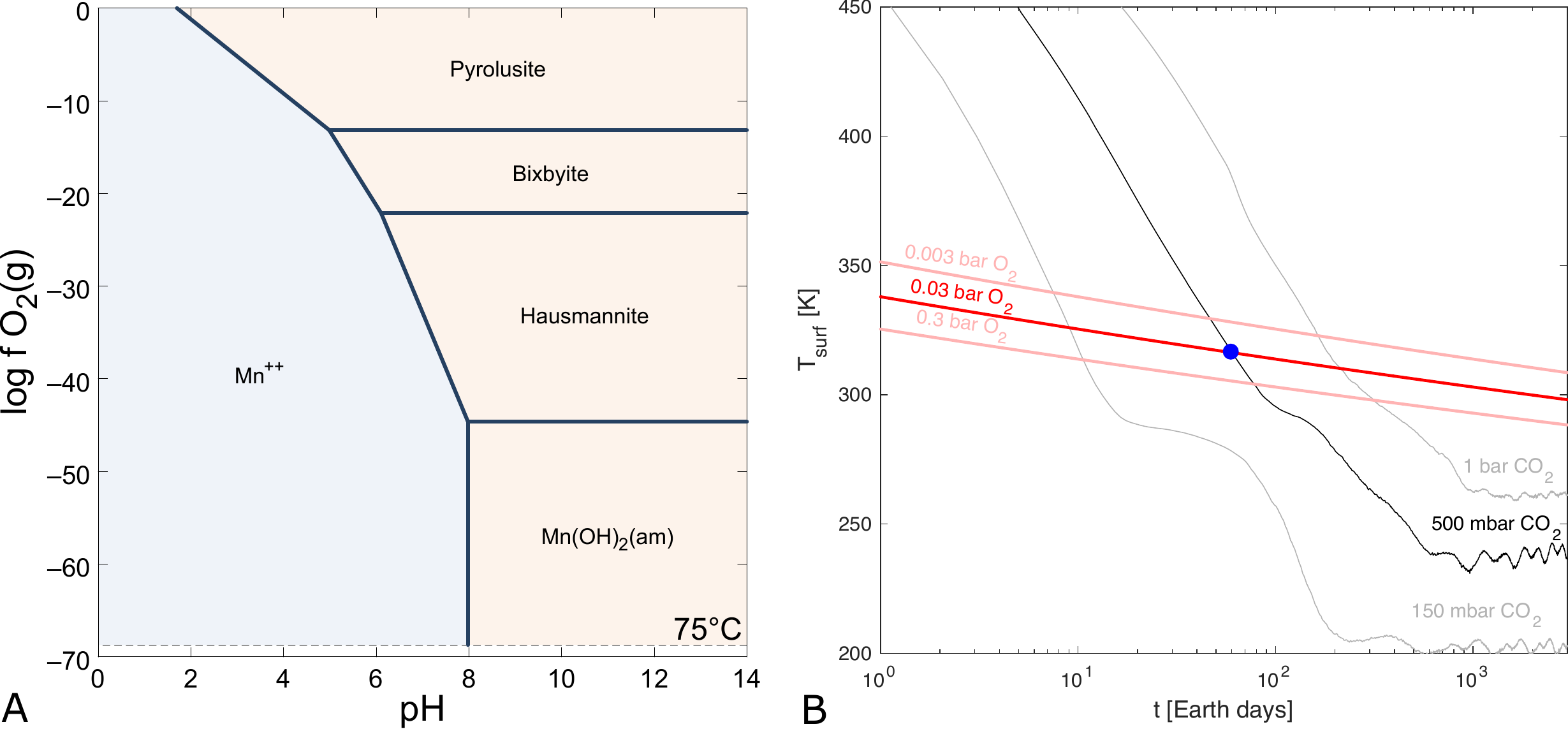}}
	\end{center}
	\caption{\textbf{Manganese oxidation on Mars during a late-stage brief warming event.} A) Equilibrium aqueous chemistry of manganese as a function of pH and oxygen fugacity, assuming a temperature of $348$~K (75~$^\circ$C) and \ce{[Mn^{2+}]} = $10^{-3}$.  B) Temperature required for a given reaction time [inverse rate constant; \cite{methods}] for manganese oxidation at a pH of 8 and various values of atmospheric partial \ce{O2} pressure (colored lines). For comparison, the surface temperature on Mars vs. time following a 30~km diameter bolide impact at three different atmospheric pressures, based on 3D climate model results, is also shown. The blue dot indicates the time at which the manganese reaction timescale equals the time passed since impact; before this point manganese oxidation is rapid, while after it the reaction becomes quenched. }
\label{fig:Mn_oxidation}
\end{figure}
\clearpage

\end{document}